\documentclass[aip,jcp,reprint,amsmath,amssymb,floatfix,citeautoscript]{revtex4-1}
\usepackage{cancel}
\usepackage{amsmath}
\usepackage{physics}
\usepackage{graphicx}
\usepackage{amssymb}
\usepackage{amsthm}
\usepackage{bm}
\usepackage{dcolumn}
\usepackage{braket}
\usepackage{longtable}

\usepackage{ragged2e}
\usepackage{txfonts}

\usepackage{color}
\usepackage[usenames,dvipsnames]{xcolor}
\definecolor{myblue}{rgb}{0,0,1}
\usepackage[breaklinks=true,colorlinks=true,linkcolor=myblue,urlcolor=myblue,citecolor=myblue]{hyperref}

\let\vr\undefined
\newcommand{\vr}{{\bm{r}}}

\begin{document}

\title{Random-phase approximation excitation energies from
approximate equation-of-motion ring coupled-cluster doubles}

\author{Timothy C. Berkelbach}
\email{berkelbach@uchicago.edu}
\affiliation{Department of Chemistry and James Franck Institute,
University of Chicago, Chicago, Illinois 60637, USA}

\begin{abstract}
The ground-state correlation energy calculated in the random-phase approximation
(RPA) is known to be identical to that calculated using a subset of terms
appearing in coupled-cluster theory with double excitations.  In particular,
this equivalence requires keeping only those terms that generate
time-independent ring diagrams, in the Goldstone sense.  Here I show that this
equivalence extends to neutral excitation energies, for which those calculated
in the RPA are identical to those calculated using an approximation to
equation-of-motion coupled-cluster theory with double excitations (EOM-CCD).  The
equivalence requires three approximations to EOM-CCD: first, the ground-state
double-excitation amplitudes are obtained from the ring-CCD equations (the same as
for the correlation energy); second, the EOM eigenvalue problem is truncated
to the single-excitation (one particle + one hole) subspace; third, the
similarity transformation of the Fock operator must be neglected, as it
corresponds to a dressing of the single-particle propagator, which is not
present in the conventional RPA.
\end{abstract}

\maketitle

\section{Introduction}

The random-phase approximation (RPA) plays a foundational role in quantum
chemistry, condensed-matter physics, materials science, and nuclear 
physics~\cite{RingSchuck,PinesNozieres,Eshuis2012}.  As
a theory of the ground-state correlation energy, the RPA is an infinite-order
resummation of all time-independent ring diagrams, which critically controls the
leading-order divergence in the energy of metals at high
density~\cite{Bohm1953,GellMann1957,PinesNozieres}.  Especially
when combined with density functional theory via the adiabatic connection
fluctuation-dissipation theorem~\cite{Langreth1975,Langreth1977} the RPA also
provides a good description of long-range dispersion
interactions~\cite{Dobson1999,Furche2001,Fuchs2002,Lebegue2010,Eshuis2012}.

The RPA correlation energy terms are a subset of those included in
coupled-cluster theory with double excitations (CCD).  Therefore, an approximate
solution of the CCD equations, known as ring-CCD, can be used to calculate the
RPA correlation energy, as shown by Freeman for the electron
gas~\cite{Freeman1977} and proven analytically by Scuseria, Henderson, and
Sorensen~\cite{Scuseria2008}; see also
Refs.~\onlinecite{Jansen2010,Scuseria2013} for subsequent studies and
generalizations.

Alternatively, the RPA may be viewed as a theory of the dynamical
polarizability, a context in which it is known to be identical to time-dependent
Hartree or Hartree-Fock~\cite{McLachlan1964}.  For finite systems, such as
molecules, the RPA leads to reasonably accurate electronic
excitations~\cite{Oddershede1978} and underlies the successful time-dependent
density functional theory~\cite{Runge1984,Stratmann1998,Hirata1999,Casida2012,Ullrich}.
For solids, the RPA polarizability correctly predicts the properties of the
collective plasmon excitation~\cite{PinesNozieres} and forms the basis for
screening the popular $GW$ approximation~\cite{Hedin1965}.  Analogous to the
correlation energy, the RPA polarizability is a resummation of all
time-dependent ring diagrams.  This similarity suggests
a relation between excitation energies calculated with the RPA and those
calculated with an approximate version of coupled-cluster theory.  In this
manuscript, I provide the precise recipe for this analogy, showing that the
RPA excitation energies (with or without exchange) can be obtained from an
approximation to electronic-excitation equation-of-motion coupled-cluster theory
with double excitations (EOM-CCD).

\section{Theory}

The dynamical polarizability is the time-ordered density-density response
function~\cite{PinesNozieres},
\begin{equation}
\Pi(\vr_1,t_1;\vr_2,t_2) = -i\langle \Psi_0 | T\left[ \delta n(\vr_1,t_1) \delta n(\vr_2,t_2) \right] | \Psi_0\rangle
\end{equation}
where $|\Psi_0\rangle$ is the ground-state wavefunction, $T$ is the time-ordering
operator, and $\delta n(\vr,t) =  n(\vr,t)-n_0(\vr)$ is the density fluctuation
away from the ground-state density.
In the frequency domain, the poles of the polarizability occur at 
all electronic excitation energies $\Omega_\nu$, with residues given by the
square of the transition densities $|\langle \Psi_0 | n(\vr) | \Psi_\nu\rangle|^2$.

In the usual diagrammatic route~\cite{KadanoffBaym}, the RPA polarizability is
expressed in terms of the irreducible polarizability $\Pi_0$ via $\Pi = \Pi_0 +
\Pi_0 [v+K] \Pi$, where $v$ is the direct Coulomb interaction and $K$ is its
exchange counterpart.  Taking the irreducible
polarizability to be simply that of a noninteracting particle-hole pair, $\Pi_0
= -i G_0 G_0$, generates the conventional RPA polarizability as a sum over all
time-dependent ring diagrams.  The location of the poles of the RPA
polarizability, i.e.~the excitation energies, are the eigenvalues of the
well-known RPA matrix, given in the following subsection.

In order to precisely relate the RPA excitation energies to those of an
approximate EOM-CCD calculation, in Sec.~\ref{ssec:rpa} I perform a
downfolding of the RPA matrix into the single particle-hole excitation subspace;
in Sec.~\ref{ssec:eom} I show that this matrix is identical to the one obtained
from EOM-CCD in the single particle-hole excitation subspace when the
ground-state double excitation amplitudes satisfy the ring-CCD equations and the
similarity transformation of the Fock operator is neglected. 
Having established the algebraic equivalence of the RPA excitation energies
and those from approximate EOM-CCD, in Sec.~\ref{ssec:diagrams} I analyze
the time-dependent Goldstone diagrams in the RPA polarizability and their
construction in the coupled-cluster framework, with special attention paid to
the non-Tamm-Dancoff diagrams; I also address the inclusion or neglect of
exchange.

\subsection{RPA excitation energies}
\label{ssec:rpa}

The RPA eigenvalue problem is given by the system of equations~\cite{RingSchuck,Scuseria2008}
(for simplicity, assuming real orbitals throughout)
\begin{equation}
\label{eq:rpa}
\left(
\begin{array}{rr}
\mathbf{A}   & \mathbf{B} \\
-\mathbf{B} & -\mathbf{A}
\end{array}
\right)
\left(
\begin{array}{c}
\mathbf{X} \\
\mathbf{Y}
\end{array}
\right)
= 
\left(
\begin{array}{c}
\mathbf{X} \\
\mathbf{Y}
\end{array}
\right) \mathbf{\Omega},
\end{equation}
where
\begin{subequations}
\begin{align}
A_{ia,jb} &= (\varepsilon_a - \varepsilon_i) \delta_{ab}\delta_{ij} + \langle ib || aj \rangle, \\ 
B_{ia,jb} &= \langle ij || ab \rangle
\end{align}
\end{subequations}
and $\mathbf{\Omega}$ is a diagonal matrix of RPA excitation energies,
which come in positive and negative pairs.
The antisymmetrized two-electron integrals are defined by 
$\langle pq||rs\rangle = \langle pq|rs\rangle - \langle pq|sr\rangle$, with
\begin{equation} 
\langle p r | rs \rangle
    = \int d\vr_1 \int d\vr_2 \phi_p(\vr_1) \phi_q(\vr_2) r_{12}^{-1} \phi_r(\vr_1) \phi_s(\vr_2),
\end{equation}
and the indices $i,j,k,l$ are used to denote occupied orbitals and $a,b,c,d$
to denote unoccupied orbitals.
Formally solving the second equation,
$-\mathbf{B}\mathbf{X} - \mathbf{A}\mathbf{Y} = \mathbf{Y} \mathbf{\Omega}$,
gives
\begin{equation}
\label{eq:y}
\mathbf{Y} = -(\mathbf{A} + \mathbf{Y}\mathbf{\Omega}\mathbf{Y}^{-1})^{-1}\mathbf{B}\mathbf{X}.
\end{equation}
Using this expression to replace $\mathbf{Y}$ in the first of the RPA equations
leads to an eigenvalue problem for $\mathbf{X}$ only,
\begin{equation}
\label{eq:rpa_downfold}
\left[\mathbf{A} - \mathbf{B}(\mathbf{A} + \mathbf{Y}\mathbf{\Omega}\mathbf{Y}^{-1})^{-1}\mathbf{B}\right]
    \mathbf{X} = \mathbf{X}\mathbf{\Omega}.
\end{equation}
Therefore, the matrix on the left-hand side, which only has support in a
single particle-hole 
excitation subspace (and not the subspace twice as large), has all
of the positive RPA excitation energies as its eigenvalues.  As written,
Eq.~(\ref{eq:rpa_downfold}) is not practical because the construction of the
downfolded matrix requires knowledge of all eigenvalues and the $\mathbf{Y}$
component of all eigenvectors; however, the matrix in
Eq.~(\ref{eq:rpa_downfold}) can be shown to be identical to an approximate
matrix derived from EOM-CCD.

\subsection{Approximate EOM-CCD}
\label{ssec:eom}

In the typical EOM-CCSD approach~\cite{Stanton1993,Krylov2008}, the $T_1$ and
$T_2$ amplitudes are obtained from the ground-state CCSD equations, and the EOM
eigenvalue problem is obtained by projecting the similarity-transformed
normal-ordered Hamiltonian, $\bar{H}_\mathrm{N} \equiv e^{-T}He^{T}
-E_\mathrm{CC}$, into a basis of singly- and doubly-excited determinants. By
contrast, to construct the relation with the RPA requires only the similarity
transformation due to $T_2$ (i.e.~$T_1 = 0$) projected only in the space of
singly-excited determinants, leading to
\begin{equation}
\label{eq:Hbar}
\langle \Phi_i^{a} | \bar{H}_\mathrm{N} | \Phi_j^b \rangle
= F_{ab} \delta_{ij} - F_{ij} \delta_{ab}
    + W_{jabi}
\end{equation}
where~\cite{Gauss1995}
\begin{subequations}
\label{eq:ints}
\begin{align}
F_{ab} &= \varepsilon_a \delta_{ab} 
    - \frac{1}{2} \sum_{klc} \langle kl||bc\rangle t_{kl}^{ac}, \\
F_{ij} &= \varepsilon_i \delta_{ij} 
    + \frac{1}{2} \sum_{kcd} \langle ik||cd\rangle t_{jk}^{cd}, \\
W_{ibaj} &= \langle ib || aj \rangle + \sum_{kc} \langle ik || ac \rangle t_{kj}^{cb}.
\end{align}
\end{subequations}
However, the similarity-transformed Fock operators lead to a dressing of the
single-particle propagators in the polarizability, in a manner which is absent
in the RPA (see Sec.~\ref{ssec:diagrams}); neglecting this effect of $T_2$ gives 
\begin{equation}
\begin{split}
\langle \Phi_i^{a} | \tilde{H}_\mathrm{N} | \Phi_j^b \rangle
&\equiv \langle \Phi_i^{a} | [f_\mathrm{N} + e^{-T_2} V_{\mathrm{N}} e^{T_2}] | \Phi_j^b \rangle \\
&= (\varepsilon_a - \varepsilon_i) \delta_{ab}\delta_{ij}
    + W_{ibaj}
\end{split}
\end{equation}
Using the definition of the $\mathbf{A}$ and $\mathbf{B}$ matrices
leads to
\begin{equation}
\label{eq:hbar}
\langle \Phi_i^{a} | \tilde{H}_\mathrm{N} | \Phi_j^b \rangle
    = A_{ia,jb} + \sum_{kc} B_{ia,kc} t_{kj}^{cb} 
    = \left[ \mathbf{A} + \mathbf{B}\mathbf{T_2} \right]_{ia,jb},
\end{equation}
where $[\mathbf{T_2}]_{ia,jb} = t_{ij}^{ab}$.
As shown in Refs.~\onlinecite{RingSchuck,Scuseria2008}, the ring-CCD equations
\begin{equation}
\begin{split}
&t_{ij}^{ab}(\varepsilon_i+\varepsilon_j-\varepsilon_a-\varepsilon_b)
  = \langle ab || ij \rangle \\
  &\hspace{1em} + \sum_{ck} t_{ik}^{ac} \langle kb || cj \rangle
  + \sum_{ck} \langle ak || ic \rangle t_{kj}^{cb}
  + \sum_{cdkl} t_{ik}^{ac}\langle kl || cd \rangle t_{lj}^{db} \label{eq:ringccd}
\end{split}
\end{equation}
can be solved in closed form in terms of the eigenvectors of the RPA equations~(\ref{eq:rpa}),
$\mathbf{T_2} = \mathbf{Y}\mathbf{X}^{-1}$.
Using this and Eq.~(\ref{eq:y}) in Eq.~(\ref{eq:hbar}) leads to the matrix
\begin{equation}
\langle \Phi_i^{a} | \tilde{H}_\mathrm{N} | \Phi_j^b \rangle
    = \left[\mathbf{A} - \mathbf{B}(\mathbf{A} + \mathbf{Y}\mathbf{\Omega}\mathbf{Y}^{-1})^{-1}\mathbf{B}\right]_{ia,jb}
\end{equation}
in exact agreement with Eq.~(\ref{eq:rpa_downfold}).
Therefore, the similarity-transformed Hamiltonian, using $T_2$ amplitudes that
satisfy the ring-CCD equations, has exactly the RPA eigenvalues when truncated to the
single-excitation subspace and transformation of the Fock operator is neglected.
Likewise, the EOM single-excitation operator $R_1 = \sum_{ai} r_i^a a_a^\dagger a_i$,
which gives the EOM-CC eigenstate,
has amplitudes that are exactly equal to the columns of $\mathbf{X}$,
i.e.~$r_i^a = X_{ia}$.

This proven equivalence can now be seen readily in the reverse direction.
Equation~(\ref{eq:hbar}) clearly implies the eigenvalue
problem $\mathbf{A}\mathbf{X} + \mathbf{B}\mathbf{T_2}\mathbf{X} = \mathbf{X}\mathbf{\Omega}$.
Using the solution of the ring-CCD equations in terms of the RPA eigenvectors, 
$\mathbf{T_2} = \mathbf{Y}\mathbf{X}^{-1}$, leads to
$\mathbf{A}\mathbf{X} + \mathbf{B}\mathbf{Y} = \mathbf{X}\mathbf{\Omega}$,
which is precisely the first of the RPA system of equations.

For the sake of discussion, I call the method described by
Eqs.~(\ref{eq:Hbar}) and (\ref{eq:ints}) EOM(S)-CCD, denoting a
CCD ground state and single-excitation EOM treatment.
When the CCD equations are approximated by the ring-CCD equations, I call
the method EOM(S)-rCCD.  Finally, when the transformation of the Fock operator
is additionally neglected, I call the method EOM(Sf)-rCCD, which is identical
to the conventional RPA.

\subsection{Diagrammatic analysis and exchange}
\label{ssec:diagrams}

The time-dependent Goldstone diagrams of the RPA polarizability are
straightforward to enumerate as all ring diagrams with all possible
time-orderings.  In order to compare with coupled-cluster theory, a diagrammatic
analysis of the coupled-cluster polarization propagator is
required~\cite{Koch1990}, along the lines of
Refs.~\onlinecite{Nooijen1992,Nooijen1993} for the one-particle Green's
function.  While a forthcoming publication~\cite{Lewis2018} will present a more
detailed analysis and numerical results, the diagrams of the coupled-cluster
polarization propagator can be analyzed by cutting the diagram after each
vertex; each connected diagram at previous time can be classified as generated by
the ground-state cluster operators, the $\Lambda$ operators, or the EOM
excitation operators.

\begin{figure}[b]
\centering
\includegraphics[scale=1.0]{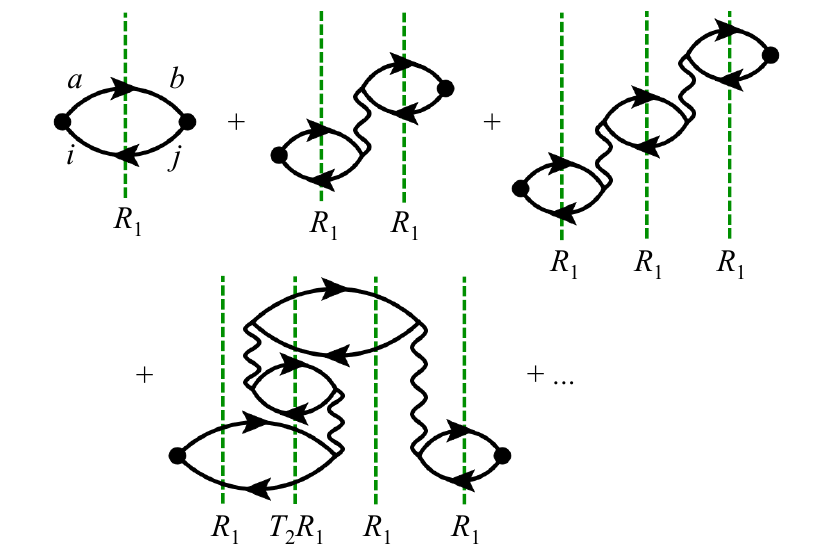}
\caption{
Time-dependent Goldstone diagrams included in the RPA polarizability,
deconstructed in terms of coupled-cluster operators $R_1$ and $T_2$.
Time increases from left to right.
}
\label{fig:rpa_pol}
\end{figure}

Figure~\ref{fig:rpa_pol} presents some example RPA ring diagrams included
through third order in perturbation theory.  Vertical cuts, indicated by dashed
lines, indicate that the first three diagrams shown are described solely by the
single-excitation EOM operator $R_1$.  These
are all examples of forward-time-ordered ring diagrams, i.e.~those resulting
from the Tamm-Dancoff approximation (TDA).  When antisymmetrized vertices are
assumed (exchange is included), then these TDA diagrams generate a
polarizability whose poles are at the excitation energies produced by
configuration interaction with single excitations.  

The fourth diagram shown in
Fig.~\ref{fig:rpa_pol} is an example of a non-TDA diagram, due to the permuted
time ordering.  Graphical analysis shows that this diagram is generated through
a combination of the $T_2$ and $R_1$ operators.  
It is straightforward to show that all non-TDA
ring diagrams included in the RPA can be deconstructed in the same manner,
using disconnected products of $T_2$ and $R_1$, but never 
the EOM $R_2$ double excitation operator; this is why
it was sufficient in Sec.~\ref{ssec:eom} to analyze the EOM eigenvalue equation
in the single-excitation subspace only.
Therefore, the RPA polarizability diagrams are exactly those produced by the
EOM(Sf)-rCCD approach.

\begin{figure}[t]
\centering
\includegraphics[scale=1.0]{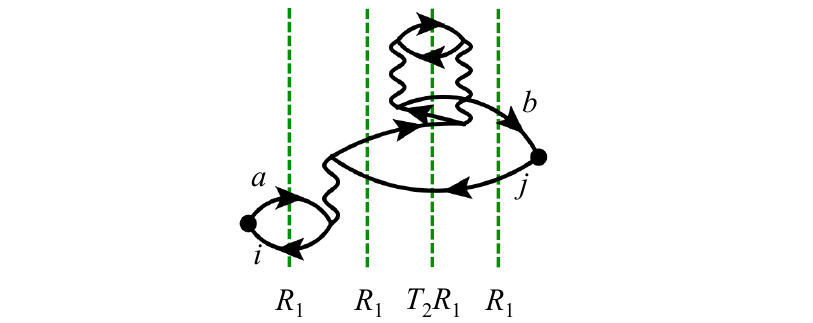}
\caption{
An example EOM(S)-rCCD polarizability diagram that is not included in the usual
RPA polarizability.  The similarity transformation of the Fock operator leads to
a dressing of the single-particle propagator. Time increases from left to right.
}
\label{fig:fock_dress}
\end{figure}

As explained in Sec.~\ref{ssec:eom}, this exact RPA equivalence requires the
neglect of the transformed Fock operator.
Figure~\ref{fig:fock_dress} shows
an example diagram generated by the EOM(S)-rCCD approximation, i.e.~without neglect
of this transformation.  Clearly, including such terms leads to a dressing of
the single-particle propagators used to construct the irreducible
polarizability.  In other words, this irreducible polarizability is of the RPA
form $\Pi_0 = -iGG$, where $G$ is most similar to the self-consistent
second-order Green's function. More accurately, this $G$ is self-consistently
determined by a second-order self-energy that only includes one out of two
possible time orderings; to include the other time ordering requires the EOM
$R_2$ excitation operator, and to include the other time ordering
self-consistently requires EOM excitation operators to all orders.

All equations, as presented above, include exchange.  Exchange can be trivially
removed by neglecting the antisymmetrization of the two-electron integrals in
the ring-CCD equations (leading to ``direct'' ring-CCD) and in the EOM
eigenvalue problem (with a factor of 2 arising from the product of two
antisymmetrized objects).  This leads to a time-dependent Hartree theory of
excitation energies, which is the more common variant of the RPA polarizability
in the condensed-matter physics literature.  Retaining exchange leads to
particle-hole ladder diagrams in addition to the ring diagrams shown in
Fig.~\ref{fig:rpa_pol}. The particle-hole ladder diagrams are required
for a description of excitonic effects in molecules or solids, and
are responsible for a reduction in the excitation energies compared to the
time-dependent Hartree theory that only includes direct ring diagrams.

\section{Conclusions and outlook}

To summarize, I have shown that the relation between the RPA and CCSD ground
states can be extended to all excited states, with a particular set of
additional approximations in the EOM-CCSD equations, dubbed EOM(Sf)-rCCD.  The
exact equivalence presented here has been verified numerically, using modified
implementations of the RPA and EOM-CCSD methodologies in the PySCF software
package~\cite{Sun2018}.

In the same way that previous work~\cite{Scuseria2008} established ground-state
CCD as the natural generalization of the RPA with correct fermionic behavior,
the present work proposes EOM(S)-CCD as the simplest fermionic theory of excited
states that contains RPA physics.  Naturally, this generalization comes with a
cost: for a single low-lying excited state, an RPA calculation scales as $N^4$,
whereas an EOM(S)-CCD calculation scales (canonically) as $N^6$.
This latter scaling is no worse than that of EOM-CCSD, which is clearly preferred
for a few low-lying excited states.
However, the cost to obtain \textit{all} excited states is $N^6$ for both RPA
and EOM(S)-CCD, to be compared to $N^{8}$ for EOM-CCSD (for all excited states
with dominant single-excitation character), which may be important for certain
spectral quantities.

In addition to providing a properly fermionic theory, the present manuscript
establishes the RPA polarizability diagrams as a strict subset of those from
EOM-CCSD.  In this sense, the CC hierarchy is a natural post-RPA route,
distinct from time-dependent density functional theory and, importantly,
systematically improvable. It is hoped that this connection will lead to
fruitful developments in the simulation of excited states, especially in the
condensed phase where RPA physics is essential.  For example, various CC-derived
polarizabilities can be used for a more accurate treatment of screening in the
$GW$ approximation, leading to a well-defined class of vertex corrections.
Similarly, a comparison of EOM-CCSD excited states to those predicted by the
$GW$+Bethe-Salpeter equation approach will provide further insight and sow
deeper connections between the condensed-matter and quantum chemistry
communities.  Work along both of these lines is currently in progress.

\section*{Acknowledgments}
I thank Alan Lewis and Bryan Lau for useful conversations and comments on this
manuscript.  This work was supported in part by startup funds from the
University of Chicago and by the Air Force Office of Scientific Research under
award number FA9550-18-1-0058.

\end{document}